# AUTOMATIC LABELING OF THE OBJECT-ORIENTED SOURCE CODE: THE LOTUS APPROACH

**Ra'Fat Al-Msie'deen**
Department of IT, Faculty of Information Technology, Mutah University, P.O. Box 7, Mutah 61710, Karak, Jordan
Email: rafatalmsiedeen@mutah.edu.jo

***ABSTRACT***: *Most of open-source software systems become available on the internet today. Thus, we need automatic methods to label software code. Software code can be labeled with a set of keywords. These keywords in this paper referred as software labels. The goal of this paper is to provide a quick view of the software code vocabulary. This paper proposes an automatic approach to document the object-oriented software by labeling its code. The approach exploits all software identifiers to label software code. The paper presents the results of study conducted on the ArgoUML and drawing shapes case studies. Results showed that all code labels were correctly identified.*

## 1. INTRODUCTION

Program understanding is the main activity for software maintenance and evolution [1]. The main idea of this paper is to reduce the amount of information mined from the software source code to provide a quick overview of the software code vocabularies.

This paper proposes a novel approach called *Lotus* to automatically provide labels for software code. Name of the approach inspired by the Lotus flower. Lotus stands for automatic Labeling of the ObjecT-oriented SoUrce code.

Several companies are facing problems with legacy software systems such as: software understanding and maintenance. The reason behind these problems is the absence of software documentation [2]. The quick development of software engineering approaches and the increasing of software system complexity has led to a large production of textual information contained in software artifacts (*e.g.,* software source code and design documents). As a result, numerous papers investigated the analysis of textual information contained in the software artifacts to support software engineering activities [3] such as: feature location [4] and software documentation [5].

To extract labels from the software source code, the Lotus approach relies on the software identifiers (*i.e.,* package, class, attribute and method). Basically, the most important words are included in the package, class, method, and attribute names [6]. The extracted labels can give a hint to the software developers about software code vocabularies.

The software code labeling process aims to analyze the legacy software to determine its main keywords, where the huge amounts of code need to be understood. To help a human expert to document the legacy software, the Lotus approach generates a set of labels represent the software vocabularies.

This paper proposes an automatic approach which extracts labels from software using its source code. Compared with the existing work that labeling software source code (*cf.* section 2), the novelty of the Lotus approach is that it exploits all software identifiers to generate software labels in an efficient way.

Lotus approach identifies the software code using static code parser. Then, the approach identifies all software identifiers. Then, Lotus splits the name of identifier into a set of keywords based on the CamelCase methods [7]. Then, Lotus returns each keyword to its stem based on the WordNet[1]. Finally, Lotus generates software labels. A sample execution of the Lotus approach is displayed on Figure 1.

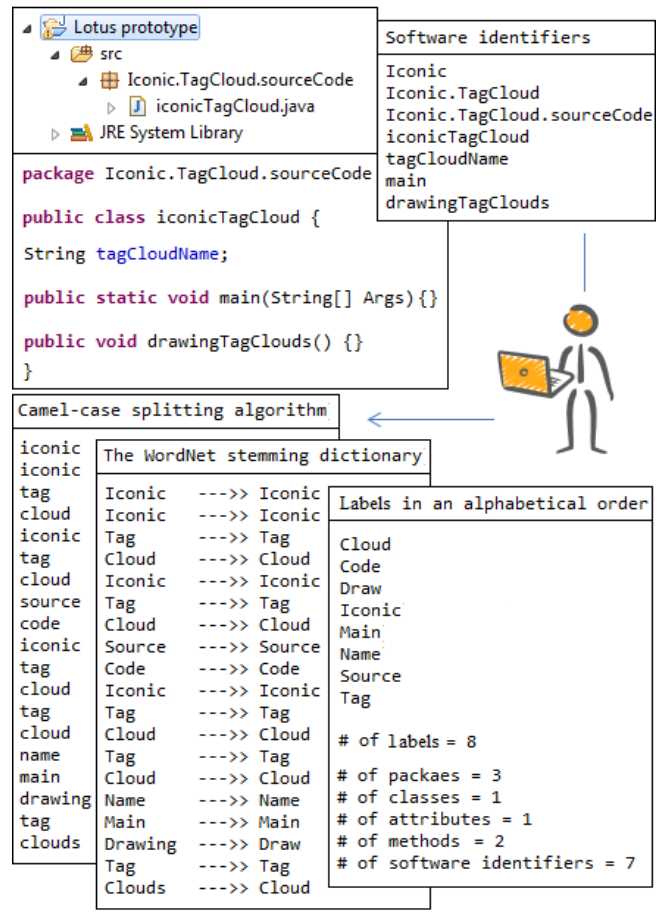

**Figure 1.** Running the Lotus approach on a simple example.

Lotus approach is detailed in the remainder of this paper as follows. Section 2 discusses the related work. Section 3 shows an overview of Lotus approach. Section 4 presents the source code labeling process step by step. Section 5 describes the experimentation, while section 6 concludes and provides perspectives for this work.

## 2. RELATED WORK

This section presents the related work and provides a concise summary of the different approaches.

Kuhn [8] presents a lexical approach that uses the log-likelihood ratios of word frequencies to automatically provide labels for software components.

AL-Msie'Deen [5] developed a tool called *Vsound* to visualize the software code and its dependencies. Vsound relies on software identifiers to visualize and document the software code. While Lotus approach extracts labels from software identifiers.

---
[1] WordNet: https://wordnet.princeton.edu/

AL-Msie'Deen *et al.* [9] proposed an approach called REVPLINE to retrieve labels for the mined features from the source code of software family based on the use-case diagrams. REVPLINE gives as output for each feature implementation, a label based on the use-case label [10].

Lucia *et al.* [6] suggest a method for source code labeling, based on information retrieval techniques, to identify relevant words in the source code. They applied numerous information retrieval methods [1] to extract terms from the names of software classes. Their approach does not consider the names of packages, attributes and methods. Lotus approach considers all granularity levels of software code.

Kuhn *et al.* [11] suggest a method to cluster software classes that use similar vocabulary together. Then, they use latent semantic indexing to automatically label clusters with their most relevant terms. Lotus approach aims to document the software code as a collection of labels.

Most existing approaches are designed to extract labels to name software features, components or clusters. In the literature, there is no work identifies code labels at all granularity levels. Also, many approaches manually extract software labels. Conversely, Lotus is designed to automatically extract software labels based on all software identifiers.

## 3. APPROACH OVERVIEW

This section provides an overview of code labeling process and describes the example that illustrates the remaining of the paper.

The main goal of Lotus approach is to understand the software code *vocabularies*. The Lotus approach aims to provide the software developers with software *labels*. The complex software system involves a huge amount of source code (*i.e.,* lines of code), so there are needs to deal with all granularity levels of source code in an abstract way.

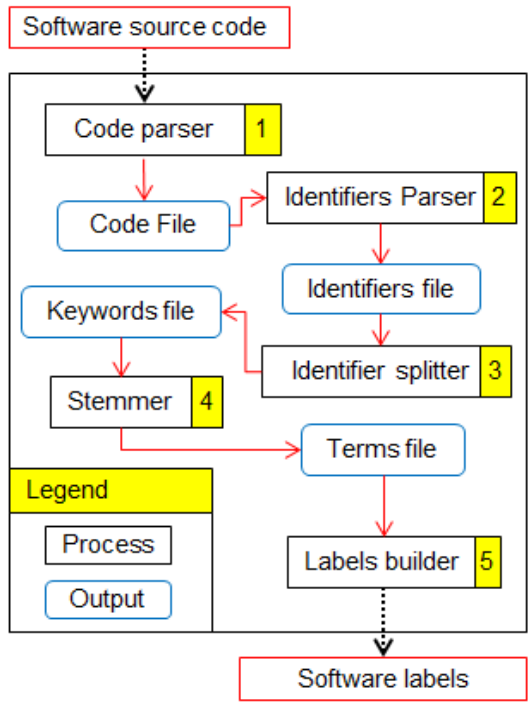

**Figure 2**. The source code labeling process.

The source code labeling process takes the software's source code as its *inputs* and creates code labels as its *outputs*. The goal of these labels is to provide us with a quick overview of software vocabularies. The *software identifier names* play a significant role in software understanding, particularly when the software documents are missing, or when the software code is very complex such that the identifier names would tell more about code to the software developers [6].

Figure 2 shows the source code labeling process in general. Lotus approach extracts software code. Then, the approach identifies software identifiers. Then, the approach splits the software identifiers into keywords based on the CamelCase method. Then, the Lotus returns each keyword to the word root via WordNet. Finally, Louts builds software labels.

As an illustrative example, Lotus considers the *drawing shapes software* [12]. The software allows user to draw three types of shapes which are *lines*, *rectangles* and *ovals*[2]. This software used to better explain the code labeling process.

## 4. SOURCE CODE LABELING PROCESS

This section describes the source code labeling process step by step. Lotus approach identifies the source code labels in *five* steps as detailed in the following.

### 4.1 Extracting source code

The Lotus approach only uses the software source code as input of the labeling process and thus Lotus does not know the code labels in advance. The first step of the labeling process aims to extract the software source code. The Lotus approach extracts all essential information from source code such as: software identifiers and code dependencies [5]. The static code parser[3] takes software code to generate the code file as output. The code file contains software identifiers. The Lotus source code parser is written completely in Java.

### 4.2 Identifying software identifiers

To comprehend the source code of legacy software systems, it's important to work with all granularity levels of the source code. Lotus approach provides us with four kinds of documents (*aka* identifiers file). Each document contains a set of identifier names. This step generates four documents which are package, class, attribute and method document.

**Table 1**. Examples of software identifiers from drawing shapes software.

| Package Names | Class Names |
|---|---|
| Drawing | MyLine |
| Drawing.Shapes | MyOval |
| Drawing.Shapes.coreElements | MyRectangle |
| Drawing.Shapes.coreFrame | PaintJPanel |
|  | DrawingShapes |

Examples of software identifiers are shown in Table 1. The package document contains the names of all software packages. The class document contains the names of all software classes and so on.

### 4.3 Splitting the identifiers

The identifier names are used to define the main entities of the software system (*e.g.,* package and class). The names of the identifiers represent a set of characters based on the rules of programming languages [13].

Lotus relies on the CamelCase technique as an identifier splitting algorithm. CamelCase method is a simple and generally used method for identifier splitting algorithms [7] and the rules of splitting are widely based on CamelCase agreement. Lotus splits identifier names into a set of keywords based on the camel-case syntax. For example, the identifier *DrawingShapes* is split into two words: drawing and Shapes.

---
[2] Drawing shapes: https://sites.google.com/site/ralmsideen/tools
[3] Static code parser: https://sites.google.com/site/ralmsideen/tools

**Table 2**. Examples of the camel-case identifier splitting algorithm.

| Identifier Name | Token/word | | |
|---|---|---|---|
| | *Word-1* | *Word-2* | *Word-3* |
| shapesArrayList | shapes | array | list |
| setCurrentColor | set | current | color |
| colorJButton | color | j | button |
| createUserInterface | create | user | interface |

Examples of camel-case splitting algorithm are shown in Table 2.

### 4.4 Stemming identifier keywords

Stemming is the procedure of stripping affixes (*i.e.,* prefixes and suffixes) from words to form a *stem* [14]. Stemming is often applied to words in information retrieval methods (*e.g.,* latent semantic indexing [12]). The stemming is used in Lotus to replace English words with their root or stem [15].

Lotus returns each keyword to its root or word stem. For instance, if we have the following keyword "*drawing*", the root of this keyword will be "*draw*". After the stemming was performed via WordNet the stems of keywords were stored in the terms file to generate the software labels.

**Table 3**. A sample of the word stems from drawing shapes software.

| Identifier word | Word stem |
|---|---|
| Pressed | Press |
| Drawing | Draw |
| Performed | Perform |
| Dragged | Drag |

Table 3 shows a sample of the word stems (*i.e.,* terms) from drawing shapes software.

### 4.5 Building the software labels

In this step, Lotus approach builds software labels to show a quick view of the used vocabularies in the software code. Figure 3 shows the mined labels from drawing shapes software.

**Figure 3**. The extracted labels from drawing shapes software.

In Lotus approach, the extracted labels called *label map*. All information about source code vocabulary given in the extracted label map. Thus, all code vocabularies can easily be acquired when needed via Lotus approach.

## 5. EXPERIMENTATION

This section presents the experiment that conducted to validate the Lotus approach. Firstly, this section presents the ArgoUML case study. Then, it presents the source code labeling outcomes and, at last, it presents the threats to validity of the Lotus approach.

In addition to the toy example used in this paper (*i.e.,* drawing shapes software), the Lotus approach has been tested on the ArgoUML[4] software system. ArgoUML is a widely used open source tool for UML modeling tool [15] and well known [16]. ArgoUML software represents a large case study (*i.e.,* 120,348 lines of code).

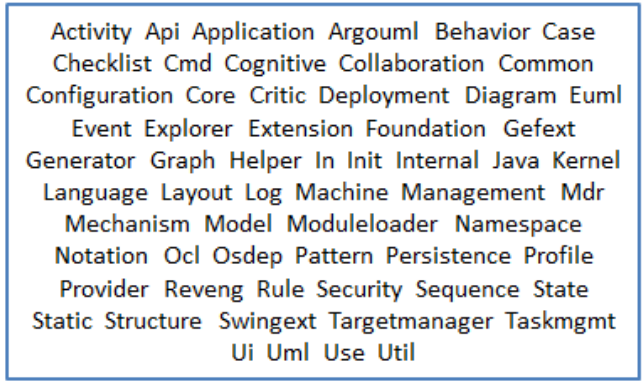

**Figure 4**. The extracted package label map from ArgoUML.

Figure 4 shows the extracted *package label map* from ArgoUML software using Lotus prototype[5]. The algorithm execution time is equal to 50860 *ms*. The identified labels (*aka* topics) are very helpful to document software vocabularies. In Figure 4, the package labels presented in an *alphabetical order* and without *repetition*.

Figure 5 shows the extracted *class label map* from ArgoUML software. The software developer can easily obtain all code labels using Lotus approach. For a lack of studies that are evaluating the extracted labels from software code, there was difficulty in evaluating the Lotus approach.

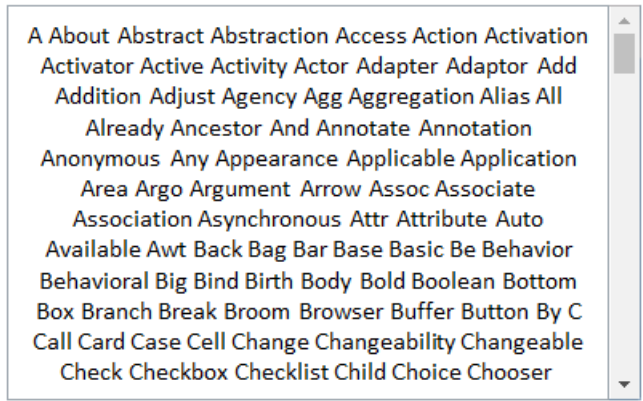

**Figure 5**. The extracted class label map from ArgoUML.

The *threat to the validity* of Lotus is that camel-case syntax may be not reliable in all cases to split software identifiers. Also, Lotus considers only the Java software, and this limits the Lotus implementation ability to deal only with Java language. Also, if the label map comes up with labels such as: *m, l, get* or *in* this does not tell a human expert much about the software system. In this case, the label map was totally useless, since the developer knowledge is missing [11].

---
[4] ArgoUML: http://argouml.tigris.org/

[5] Lotus prototype: https://sites.google.com/site/ralmsideen/tools

**Table 4**: Summary of Lotus approach.

| Objectives | | Programmed method | |
|---|---|---|---|
| Code labeling | √ | Automatic | √ |
| Code understanding | √ | Input | |
| Code documentation | √ | Software packages | √ |
| Code visualization | √ | Software classes | √ |
| Tool support | | Software attributes | √ |
| WordNet | √ | Software methods | √ |
| Evaluation | | Technique | |
| Evaluation | x | Ad hoc algorithm | √ |
| Case study | | Camel-case algorithm | √ |
| Drawing shapes | √ | Splitting method | |
| ArgoUML | √ | Camel-case | √ |
| Output | | | |
| Package label map | √ | | |
| Class label map | √ | | |
| Attribute label map | √ | | |
| Method label map | √ | | |

Table 4 summarizes Lotus approach where it presents the *objectives*, *programmed method*, *input*, *tool support*, *evaluation*, *technique*, *case study*, *splitting method* and *output*.

## 6. CONCLUSION AND FUTURE WORK

This paper proposed an approach to extract code labels. The goal of this paper is to document code vocabularies. The novelty of this paper is that it exploits all software identifiers to document software vocabularies. Lotus has implemented on several case studies. Results showed that all label maps were identified. Regarding future work, Lotus plans to use code *comments* in the code labeling process. Also, Lotus approach plans to exploit the tag cloud visualization technique [17] to display label frequency across software code. In addition, Lotus plans to study the evolution of software labels in a set of software variants using formal concept analysis [18].